\documentclass[twocolumn,showpacs,pre]{revtex4}

\usepackage{graphicx}
\usepackage{dcolumn}
\usepackage{bm}
\usepackage{amsmath}
\usepackage{graphics}
\begin{document}

\title{Long range polarization attraction between two
different likely charged macroions}

\author{Rui Zhang}
\author{B. I. Shklovskii}
\affiliation{Theoretical Physics Institute, University of
Minnesota, Minneapolis, Minnesota 55455}

\date{\today}

\begin{abstract}
It is known that in a water solution with multivalent counterions
($Z$-ions) two likely charged macroions can attract each other due
to correlations of $Z$-ions adsorbed on their surfaces. This
``correlation" attraction is short-ranged and decays exponentially
with increasing distance between macroions at characteristic
distance $A/2\pi$, where $A$ is the average distance between
$Z$-ions on the surfaces of macroions. In this work, we show that
an additional long range ``polarization" attraction exists when
the bare surface charge densities of the two macroions have the
same sign, but are different in absolute values. The key idea is
that with adsorbed $Z$-ions, two insulating macroions can be
considered as conductors with fixed but different electric
potentials. Each potential is determined by the difference between
the entropic bulk chemical potential of a $Z$-ion and its
correlation chemical potential at the surface of the macroion
determined by its bare surface charge density. When the two
macroions are close enough, they get polarized in such a way that
their adjacent spots form a charged capacitor, which leads to
attraction. In a salt free solution this polarization attractive
force is long ranged: it decays as a power of the distance between
the surfaces of two macroions, $d$. The polarization force decays
slower than the van der Waals attraction and therefore is much
larger than it in a large range of distances. In the presence of
large amount of monovalent salt, the polarization attraction
decays exponentially at $d$ larger than the Debye-H\"{u}ckel
screening radius $r_s$. Still, when $A/2\pi\ll d\ll r_s$, this
force is much stronger than the van der Waals attraction and the
correlation attraction mentioned above. The recent atomic force
experiment has shown some evidence for this polarization
attraction.
\end{abstract}

\pacs{61.20.Qg,61.25.Hq,82.45.Gj}

\maketitle

\section{Introduction}\label{sec:introduction}
Water solutions of strongly charged colloidal particles
(macroions) with multivalent ($Z$ valent) counterions ($Z$-ions)
are important in physics, chemistry, biology and chemical
engineering. Colloidal particles, charged lipid membranes, DNA,
actin, and even cells and viruses are examples of different
macroions. Multivalent metallic ions, dendrimers, charged
micelles, short DNA and other relatively short polyelectrolytes
like spermine can play the role of $Z$-ions. Several interesting,
counterintuitive phenomena have been discovered in such a system,
such as charge inversion, which have attracted significant
attention (see review paper Ref.~\cite{Nguyen-review} and
references therein). Charge inversion happens in a water solution
when a macroion binds so many $Z$-ions that its net charge changes
sign. This phenomenon is especially important for gene therapy. To
deliver DNA into the cell, charge of negative DNA (playing the
role of macroion) should be inverted by positive $Z$-ions to
approach a negatively charged cell membrane. Here we actually
assume that the membrane is weakly charged and therefore its
charge is not inverted. The question is what happens if the charge
of the membrane is also inverted by positive $Z$-ions. Can DNA
still be attracted to the membrane?

\begin{figure}[ht]
\begin{center}
\includegraphics[width=0.5\textwidth]{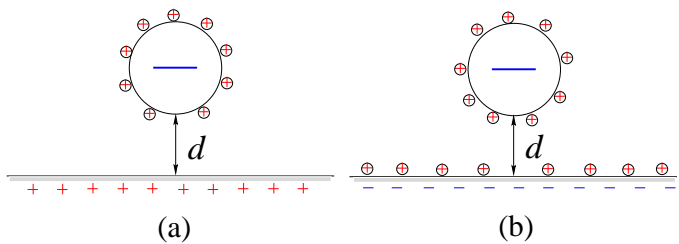}
\end{center}
\caption{Schematic illustration of two experimental setups for
atomic force measurements of likely charged macroions. The force
between the probe (the big sphere with negative charge) and the
surface is measured at different concentrations of $Z$-ions (small
spheres with positive charges). (a) In the setup of
Ref.~\cite{Lemay}, the bare surface and the bare probe are the
oppositely charged so that positive $Z$-ions are only adsorbed to
the probe and invert its charge. (b) In the setup of
Ref.~\cite{Sivan}, the bare surface and the bare probe are likely
charged so that positive $Z$-ions are adsorbed to both of them,
again making them likely charged at large $d$.}\label{fig:afm}
\end{figure}

A similar question was put forward by the recent atomic force
experiment\cite{Lemay} designed to verify the theory of charge
inversion based on strong correlations between
$Z$-ions~\cite{Boris},\cite{Nguyen-review} In the experiment,
forces between a negatively charged spherical macroion attached to
the cantilever (probe) and a positively charged surface were
measured at different concentrations of positive $Z$-ions (see
Fig.~\ref{fig:afm}a)~\cite{Lemay}. At small concentrations of
$Z$-ions, the probe is attracted to the surface. With increasing
concentration of $Z$-ions, the charge of the probe gets inverted
by $Z$-ions, and the measured force at large $d$ becomes
repulsive, where $d$ is the distance of closest proximity from the
probe to the surface. The critical concentration of $Z$-ions,
where this happens is in reasonable agreement with the prediction
of Ref.~\cite{Boris}. However, an interesting new feature of the
repulsive force was observed. With decreasing $d$, the repulsive
force reaches a maximum at $d\simeq 100$ $\AA$ (which is roughly
equal to the Debye-H\"{u}ckel screening radius of the solution)
and start to decrease. This suggests the existence of a competing
attraction.

One can also consider a different experimental setup~\cite{Sivan}
in which the probe and the surface are likely charged and both
adsorb $Z$-ions (Fig.~\ref{fig:afm}b). When the concentration of
$Z$-ions is high enough so that the charges of the surface and the
probe are both inverted at large $d$, it is interesting to find
out whether the force may be attractive. Preliminary experiments
showed such attraction ~\cite{Sivan}. Notice that this setup
brings us back to the original question of attraction of
charge-inverted DNA to charge-inverted cell membrane.

Motivated by these questions, in this paper we study the
interaction between two different macroions in the presence of a
large concentration of $Z$-ions. In the main part of the paper, we
focus on the case of two bare negatively charged macroions and
positive $Z$-ions corresponding to Fig.~\ref{fig:afm}b, since it
is pedagogically simple. Only in Sec.~\ref{sec:experiment}, we
discuss the case of oppositely bare charged macroions
(Fig.~\ref{fig:afm}a) in the connection with Ref.~\cite{Lemay}. We
assume that the valence of $Z$-ions, $Z\gg 1$, but still many
$Z$-ions are needed to neutralize one macroion. The two macroions
are spherical but different in their bare surface charge
densities. We will see that this is crucial to produce the
attraction. Actually this is the reason why this attraction was
not reported before, when the focus was on identical
macroions~\cite{Nguyen-review}. In calculation, we focus on the
two limiting cases, $R_1=R_2$ and $R_1\ll R_2$, where $R_1$ and
$R_2$ are the radii of the two macroions.

\begin{figure}[ht]
\begin{center}
\includegraphics[width=0.2\textwidth]{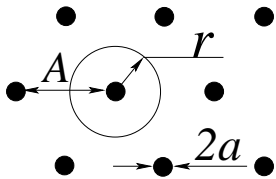}
\end{center}
\caption{$Z$-ions are strongly correlated on the surface of the
macroion. Their short range order is similar to that of a Wigner
crystal with lattice constant $A$.}\label{fig:wc}
\end{figure}

Before we discuss the mechanism of this attraction, let us first
briefly review the theory of charge inversion~\cite{Boris}. Let us
consider a water solution with one negatively charged macroion and
many positively charged $Z$-ions. Due to the Coulomb interaction,
$Z$-ions are adsorbed to the surface of the macroion. On the
surface, they strongly repel each other with energy much larger
than $k_BT$ and form a two dimensional strongly correlated liquid
with short range order similar to a Wigner crystal (WC)
(Fig.~\ref{fig:wc}). When a new $Z$-ion approaches the macroion,
it repels already adsorbed $Z$-ions and creates a negative spot
for itself. One can view it as an electrostatic image of the
$Z$-ion, similar to the image on a conventional conducting
surface. Attraction to the image leads to an additional negative
chemical potential $\mu_c$ ($c$ stands for ``correlation") for
each $Z$-ion on the surface of the macroion. As a result, when the
concentration of $Z$-ions in the solution is large enough, so many
$Z$-ions are adsorbed that the net charge of the macroion is
inverted and becomes positive~\cite{Boris}.

As pointed out in Ref.~\cite{Boris}, for an already neutral
macroion, adsorption of additional $Z$-ions can be viewed as
charging a conductor of certain capacitance with a fixed electric
potential $\phi$. This potential is determined by the difference
between the chemical potential of $Z$-ions in the solution and the
correlation chemical potential $\mu_c$ on the surface. It is a
constant all over the macroion surface because the net surface
charge density of the macroion changes very little and therefore
$\mu_c$ is a constant (see Sec.~\ref{sec:attraction} for detail
discussion). Actually, this electric potential is similar to the
surface potential of a colloidal particle determined by its
``potential-determining ions" given by the Nernst
equation~\cite{Colloid}. The only difference is that in the
present case, $\mu_c$ originates from correlations of $Z$-ions
instead of chemical bonding.

The model of conductors is convenient to discuss the interaction
and understand the possible attraction between two different
macroions in the presence of $Z$-ions. Actually, as we know from
electrostatics, if two conductors are charged with different
potentials, they may attract each other, even though the two
potentials have the same sign. For example, consider two
conducting spheres 1 and 2 with same size and potentials $\phi_1$
and $\phi_2$ respectively and $\phi_1>\phi_2>0$. When the two
spheres are far away, they repel each other. When we bring them
closer, adjacent spots of two spheres form a charged capacitor
with potential difference $\phi_1-\phi_2$. In other words, spheres
get polarized. The polarization attraction competes with the
overall repulsion between two spheres (since $q_1>q_2>0$) and
dominates at small distances between spheres. Similar polarization
leads to attraction of a conducting sphere to a close conducting
plate holding at different potentials.

Using the model of conductors, we find that as long as the bare
surface charge densities of the two macroions are different, they
always attract each other at small enough distances, even when
their bare charges or net charges are of the same sign. This
polarization attraction is even more appreciable in the case that
the sizes of the two macroions are very different as in
Fig.~\ref{fig:afm}. We also find that the attractive polarization
force $F_p$ decays with increasing distance as a power law. In the
case of Fig.~\ref{fig:afm}, at $d\ll r_s$, we get
\begin{equation}
F_p=-\frac{\pi DR_1}{d}(\phi_1-\phi_2)^2,\label{Fdrs}
\end{equation}
where $r_s$ is the Debye-H\"{u}ckel screening radius, $D=80$ is
the dielectric constant of water and $R_1$ is the radius of the
spherical probe.

We emphasize that the polarization attraction discussed here is
different from two standard attractive forces. The first one is
the well known van der Waals attraction used in standard DLVO
theory~\cite{Colloid}. In the case of Fig.~\ref{fig:afm}, it is
given by
\begin{equation}
F_{vdW}=-\frac{R_1H}{6d^2}, \label{FvdW}
\end{equation}
where $H$ is the Hamaker constant. It is clear that $F_{vdW}$
decays with $d$ faster than $F_p$. The ratio of these two forces
is
\begin{equation}
\frac{F_p}{F_{vdW}}=\frac{6\pi
Dd}{H}(\phi_1-\phi_2)^2.\label{ratio}
\end{equation}
For the typical $H=1.0\times 10^{-20}J=2.4k_BT$ and reasonable
$|\phi_1-\phi_2|\simeq 0.4k_BT/e$ (see Sec.~\ref{sec:attraction}),
this ratio is larger than unity if $d>6$ $\AA$.

The second competing force is the short range attraction between
two macroions due to correlations of $Z$-ions on their
surfaces~\cite{Bloomfield}. In the spot where two macroions touch
each other, the surface density of $Z$-ions is doubled and the
correlation energy is gained. This correlation attraction is
related to Wigner-crystal-like arrangement of $Z$-ions on the
surfaces and therefore decays exponentially with increasing
distance as $e^{-2\pi d/A}$, where $A$ is the ``lattice constant"
of the Wigner crystal (see Fig.~\ref{fig:wc}). For $d\gg A/2\pi$,
it becomes much weaker than the polarization attraction given by
Eq.~(\ref{Fdrs}) (we assume that $r_s\gg A/2\pi$). In this sense
the polarization attraction is a \emph{long range} force.

One can understand the polarization attraction from another point
of view, i.e., from the concept of contact electrification. As
well known, when two different solids contact to each other in
vacuum, due to the difference in their work functions, certain
amount of electrons move from one material to the other,
developing a electric voltage which stops further charging. This
contact electrification leads to the well known Coulomb attraction
between them~\cite{Landau}. Here a contact between two objects is
necessary to eliminate the kinetic barrier for electrons and makes
electrification possible during time of experiments. If we wait
long enough, the electrification can happen even without a close
contact. In our case of macroions in water, $Z$-ions play the role
of electrons and the absolute value of chemical potential $\mu_c$
of a $Z$-ion on the surface of the macroion plays the role of the
work function. To keep the same electro-chemical potential of
$Z$-ions, since $\mu_c$ is different for two macroions, the
electric potentials must also be different, which leads to
attraction. The difference in our system is that the kinetic
barrier for $Z$-ions is relatively small and the equilibrium of
$Z$-ions is easily achieved through the solution.

Similar electrification phenomenon has been studied in
Ref.~\cite{Messina} in a toy model of two negative spherical
macroions exactly neutralized by $Z$-ions. The two spheres had the
same radius, but substantially different bare charges. It was
shown that under conditions of total neutrality one sphere becomes
undercharged (negative) and the other is overcharged (positive)
and, therefore, they attract each other.

In the present paper, we discuss a more generic and realistic
situation: there is certain concentration of $Z$-ions in the
solution (a fixed chemical potential of Z-ions) and the macroions
can be both either undercharged or overcharged by them. In spite
of the same sign of their net charges these macroions attract each
other at small distances because of local polarization around the
points of closest proximity.

In the presence of large amount of monovalent salt, the Coulomb
interaction is effectively truncated at the Debye-H\"{u}ckel
screening radius $r_s$. When $\phi_1=\phi_2$, the two macroions
with adsorbed $Z$-ions repel each other as described by the
standard DLVO theory~\cite{Colloid}. However, when
$\phi_1\neq\phi_2$, i.e., when the two macroions are different,
the polarization attraction appears. When $\phi_1$ and $\phi_2$
have the same sign, it decays as $e^{-2d/r_s}$ at distances larger
than $r_s$. Still, for $A/\pi \ll d \ll r_s$, this attraction is
much stronger than both the van der Waals
attraction~\cite{Colloid} and the correlation
attraction~\cite{Bloomfield}, which is proportional to $e^{-2\pi
d/A}$.

This paper is organized as follows. In Sec.~\ref{sec:attraction},
we describe the interaction between two different macroions in the
presence of $Z$-ions and show that it is similar to the
interaction between two conductors with different potentials. In
Sec.~\ref{sec:metal}, we discuss the interaction of two conducting
spheres and derive the power law of the attractive force. In
Sec.~\ref{sec:screening}, we take into account of the effect of
screening by monovalent salt. In Sec.~\ref{sec:experiment}, we
generalize our theory to the case when only one macroion adsorbs
$Z$-ions in the connection with the experiment in
Ref.~\cite{Lemay} (see Fig.~\ref{fig:afm}a). We conclude in
Sec.~\ref{sec:conclusion}.

\section{The model of conductors for two macroions in the presence
of $Z$-ions}
\label{sec:attraction}

In this section we show that the interaction between two macroions
in the presence of $Z$-ions can be considered as the interaction
between conductors with different electric potentials. Let us
first consider a single spherical macroion with radius $R$ and
\emph{bare} charge $-Q<0$ in a water solution with $Z$-ion
concentration $n$. The bare surface charge density of the macroion
is $-\sigma=-Q/4\pi R^2$. As we discussed in
Sec.~\ref{sec:introduction}, $Z$-ions on the surface of the
macroion strongly repel each other to form a strong correlated
liquid, similar to a structure of Wigner crystal in the short
range (see Fig.~\ref{fig:wc}). The chemical potential related to
this correlation is dominated by its low temperature expression,
which can be estimated from the Coulomb interaction inside each WC
cell (see Eq.~(9) in Ref.~\cite{Boris} for the full finite
temperature expression of $\mu_c$)
\begin{equation}
\mu_c(s)\simeq-\frac{1.65Z^2e^2}{Dr}=-\frac{1.65\sqrt{\pi
s}Z^2e^2}{D}.\label{muc}
\end{equation}
Here $s$ is the surface density of $Z$-ions on the macroion, and
$r=A\sqrt{\sqrt{3}/2\pi}$ is the radius of the WC cell (see
Fig.~\ref{fig:wc}). It satisfies $\pi r^2s=1$.

The equilibrium condition of a $Z$-ion in the solution then reads
\begin{equation}
\frac{ZeQ^{\ast}}{DR}+\mu_c(s)
=-k_BT\ln\left(\frac{s^{3/2}}{n}\right),\label{eqcon}
\end{equation}
where $Q^{\ast}$ is the \emph{net} charge of the macroion combined
with adsorbed $Z$-ions. The left hand side is the electrostatic
energy of a $Z$-ion at the surface of the macroion plus the
correlation chemical potential. The right hand side is the
difference between the entropic parts of the chemical potentials
of a $Z$-ion in the solution and at the surface. As we will see
below, this equilibrium condition is the key for the analogy with
conductors.

When $n$ increases, the number of $Z$-ions adsorbed to the
macroion also increases. At some critical $n=n_0$, $Q^{\ast}$
becomes zero. According to Eq.~(\ref{eqcon}), we have
\begin{equation}
n_0=s_0^{3/2}\exp\left(\frac{-|\mu_c(s_0)|}{k_BT}\right).\label{n0}
\end{equation}
Here $s_0=\sigma/Ze$ is the surface density of $Z$-ions to
neutralize the macroion. The corresponding radius of WC cell is
$r_0$. In the case we are interested, $\mu(s_0)\simeq
Z^2e^2/Dr_0\gg k_BT$ (this defines ``strongly charged" $Z$-ions).
Therefore $n_0$ is an exponentially small concentration which is
easy to reach in experiments. For $n>n_0$, more $Z$-ions come to
the macroion and $Q^{\ast}$ becomes positive. According to
Eq.~(\ref{eqcon}), when $n$ is so large that the entropy term is
completely negligible, the maximum value of inverted $Q^{\ast}$,
$Q_{max}^{\ast}$, is achieved~\cite{Boris},
\begin{equation}
Q^{\ast}_{max}=0.8\sqrt{QZe}.\label{Qmax}
\end{equation}
It is calculated by assuming
\begin{equation}
s\simeq s_0,\ \ \mu_c(s)\simeq \mu_c(s_0).\label{muc0}
\end{equation}
This assumption is self-consistent, since for $Q\gg Ze$ we have
$(s_{max}-s_0)/s_0=Q_{max}^{\ast}/Q=0.8\sqrt{Ze/Q}\ll 1$.

Now we can clearly see the similarity between a neutralized
macroion and a neutral conductor. For a given $n>n_0$, the
charging process from $Q^{\ast}=0$ to $Q^{\ast}>0$ can be viewed
as charging a conductor with fixed potential. Indeed, the
equilibrium condition Eq.~(\ref{eqcon}) can be written as
\begin{equation}
Q^{\ast}=DR\phi,\label{Qasts}
\end{equation}
with
\begin{equation}
\phi=\frac{|\mu_c(s_0)|-k_BT\ln(s_0^{3/2}/n)}{Ze}.\label{phi}
\end{equation}
Notice that $\phi$ can be expressed through $s_0$ because of
Eq.~(\ref{muc0}). Eq.~(\ref{Qasts}) is exactly the same as the
expression of charge for a conductor with capacitance $C=DR$ and
charged at fixed potential $\phi$. For $n<n_0$, by similar
discussion, we have a conductor charged at $\phi<0$, providing
that $|Q^{\ast}(n)|\ll Q$ so that Eq.~(\ref{muc0}) is still
satisfied.

Now let us consider two macroions in a water solution with
$Z$-ions. Clearly, when $d$ is very large, if $n$ is close to
$n_0$ for both macroions, the analogy to conductors holds. The two
macroions have electric potentials $\phi_1$ and $\phi_2$. When $d$
decreases, due to the interaction between two macroions, $s$
changes and becomes nonuniform on their surfaces. When $d$ is very
small, $s$ may become substantially different from $s_0$ at
closest spots of the two surfaces. Then Eq.~(\ref{muc0}) is not
valid and $\phi_{1,2}$ start to change with $d$~\cite{potential}.
In this paper, we stay in the limit of large $s_0$ so that
$|s-s_0|\ll s_0$. Accordingly, Eq.~(\ref{muc0}) is always valid
and our approximation of fixed potentials is fine.

Rewriting Eq.~(\ref{phi}) for two macroions, we have
\begin{equation}
\phi_{1,2}=\frac{|\mu_c(s_{1,2})|-k_BT\ln(s_{1,2}^{3/2}/n)}{Ze}.
\label{phi12}
\end{equation}
Here the subindexes $1,2$ represent two macroions respectively.
$s_{1,2}$ are the number densities of $Z$-ions which neutralize
two macroions, i.e., $s_{1,2}=\sigma_{1,2}/Ze$. Notice that
$\phi_{1,2}$ are completely determined by $n$ and $s_{1,2}$, i.e.,
the concentration of $Z$-ions and the bare surface charge
densities of the two macroions. For a given solution of $Z$-ion
concentration $n$, if $\sigma_1\neq \sigma_2$, $\phi_1\neq\phi_2$.
For example, for typical values $Z=+3$, $n=1$ mM,
$-\sigma_1=-0.75$ e/nm$^2$ and $-\sigma_2=-0.45$ e/nm$^2$, we have
$\phi_1=0.77 k_BT/e$, $\phi_2=0.37k_BT/e$ and
$|\phi_1-\phi_2|=0.40k_BT/e$. Here $\mu_c(s_1)=-7.6k_BT$ and
$\mu_c(s_2)=-5.7k_BT$ are calculated using the full expression of
$\mu_c$ given by Eq.~(9) in Ref.~\cite{Boris}.

In order to calculate the force between two macroions, we can
imagine that they are conductors in ion-free water with potentials
$\phi_1$ and $\phi_2$ supported by two batteries. Such conductors
are not in equilibrium with each other while our macroions are in
equilibrium. This however does not matter for the calculation of
the force which is the same in both cases. Indeed, the force
depends only on potentials and capacitance matrix of the system.

\section{Attraction of two conducting spheres
with different potentials} \label{sec:metal}

\begin{figure}[ht]
\begin{center}
\includegraphics[width=0.45\textwidth]{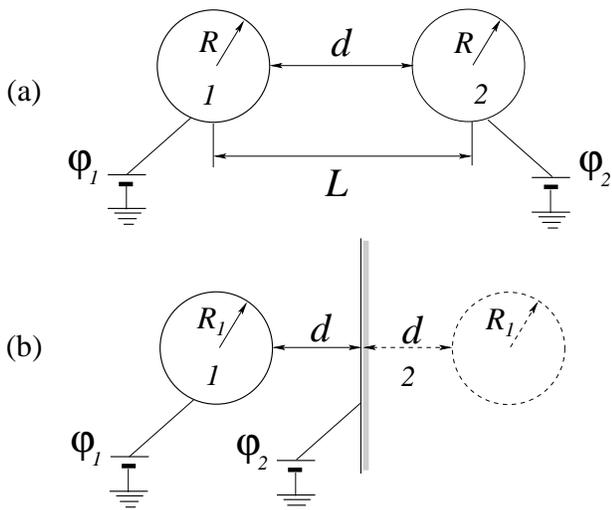}
\end{center}
\caption{The interaction between two spherical macroions in the
presence of $Z$-ions can be considered as the interaction between
two conducting spheres with fixed but different electric
potentials. We focus on two cases: (a) The size of the two spheres
are equal. (b) The size of sphere 2 is much larger than sphere 1
(the radius of sphere 2, $R_2$, is not shown). For $d\ll R_2$, the
interaction between two spheres can be considered as the
interaction between sphere 1 with its ``image sphere" inside
sphere 2.}\label{fig:spheres}
\end{figure}

In this section, let us focus on the interaction of two conducting
spheres 1 and 2 with radii $R_1$ and $R_2$ and potentials $\phi_1$
and $\phi_2$ in ion-free water (see Fig.~\ref{fig:spheres}). As we
explained in the previous section, this interaction is equivalent
to the interaction between macroions covered by $Z$-ions. We
define the the distance of closest proximity between the two
spheres as $d$ and the distance between the centers of the two
spheres as $L=d+R_1+R_2$ (see Fig.~\ref{fig:spheres}). Below we
use $L$ or $d$ alternatively for convenience. We focus on two
limiting cases when $R_1=R_2$ and $R_1\ll R_2$, but our approach
is applicable for any two spheres.

We start from the total free energy of two conductors with fixed
potentials~\cite{Landau}
\begin{equation}
U(L)=-\frac{1}{2}C_{11}\phi_1^2-\frac{1}{2}C_{22}\phi_2^2
-C_{12}\phi_1\phi_2. \label{U}
\end{equation}
Here $C_{11}, C_{22}, C_{12}$ are the self and mutual capacitances
of the two spheres depending on $L$. Notice that in the case of
fixed potentials, the work done by the environment to charge the
two conductors should be included in the total energy and
therefore each term  in Eq.~(\ref{U}) has a negative sign. The
generic formula of the force between the two conductors is given
by
\begin{equation}
F(L)=-\frac{\partial U}{\partial L}=\frac{\partial
C_{11}}{\partial L}\frac{\phi_1^2}{2}+\frac{\partial
C_{22}}{\partial L}\frac{\phi_2^2}{2}+\frac{\partial
C_{12}}{\partial L}\phi_1\phi_2. \label{F}
\end{equation}
As will be shown, the force is always attractive when
$\phi_1\neq\phi_2$ and the two spheres are close enough. And it
decays as a power law with increasing distance between spheres.

\subsection{Two spheres of the same size}
Let us start from a simple case when the radii of the two spheres
are equal, $R_1=R_2=R$ (Fig.~\ref{fig:spheres}a). In this case,
$C_{11}=C_{22}$ and Eq.~(\ref{F}) is simplified to
\begin{equation}
F(L)=\frac{\partial C_{11}}{\partial
L}\frac{\phi_1^2+\phi_2^2}{2}+\frac{\partial C_{12}}{\partial
L}\phi_1\phi_2. \label{Fs}
\end{equation}
According to the standard method of image charges, $C_{11}$ and
$C_{12}$ can be calculated by considering a infinite series of
images induced in the two spheres~\cite{Smythe}. We have
\begin{subequations}
\label{c}
\begin{eqnarray}
C_{11}&=&DR\sinh t\sum_{n=1}^{\infty}\frac{1}{\sinh (2n-1)t},\\
C_{12}&=&-DR\sinh t \sum_{n=1}^{\infty}\frac{1}{\sinh 2nt},
\end{eqnarray}
\end{subequations}
where $t=\textrm{arccosh}(L/2R)$. Consequently,
\begin{subequations}
\label{pc}
\begin{eqnarray}
\frac{\partial C_{11}}{\partial L}&=&-\frac{D}{2\sinh t}
\sum_{n=1}^{\infty}\frac{a_{2n-1}(t)}
{\sinh^2 (2n-1)t},\label{pc1}\\
\frac{\partial C_{12}}{\partial L}&=&\frac{D}{2\sinh t}
\sum_{n=1}^{\infty}\frac{a_{2n}(t)} {\sinh^2 2nt}.\label{pc2}
\end{eqnarray}
\end{subequations}
Here $a_n(t)$ is defined as
\begin{equation}
a_n(t)=n\sinh t\cosh nt-\cosh t\sinh nt.\label{an}
\end{equation}
Since $a_n(t)>0$ for $t>0$, we have $\partial C_{11}/\partial
L<0$, $\partial C_{12}/\partial L>0$. Also one can easily prove
that $|\partial C_{11}/\partial L|<|\partial C_{12}/\partial L|$.
These inequalities lead to some simple results. Firstly, when the
signs of $\phi_1$ and $\phi_2$ are opposite, $F<0$, i.e., the
force is attractive, as it should be. Secondly, when $\phi_1=0$ or
$\phi_2=0$, we have $F<0$. Physically, the sphere with nonzero
potential induces opposite charge in the sphere with zero
potential, therefore they attract each other. Finally, when
$\phi_1=\phi_2$, $F>0$, i.e., the force is repulsive. Indeed, the
two spheres with same potential can be considered as a capacitor
as a whole. Obviously, the closer the two spheres are, the smaller
the capacitance is, and the higher the energy is (remember the
negative sign in Eq.~(\ref{U})).

\begin{figure}[ht]
\begin{center}
\includegraphics[width=0.45\textwidth]{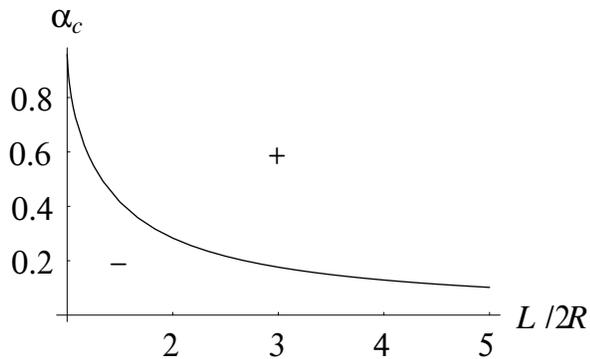}
\end{center}
\caption{The critical $\alpha_c=\phi_2/\phi_1$, at which the
interactive force between two spheres is zero, is plotted as a
function of $L/2R$ numerically. The minus (plus) sign under
(above) the curve means that the interaction between the two
spheres is attractive (repulsive).}\label{fig:alpha}
\end{figure}

We are specifically interested in the case where the signs of
$\phi_1$ and $\phi_2$ are the same but the magnitudes are
different. Without losing generality, we consider the case of
$\phi_1>\phi_2>0$. Introducing $\alpha=\phi_2/\phi_1$, we have
\begin{equation}
F(L)=\left[\alpha-\frac{(\alpha^2+1)}{2}\left|\frac{\partial
C_{11}/\partial L}{\partial C_{12}/\partial L}\right|
\right]\left|\frac{\partial C_{12}}{\partial
L}\right|\phi_1^2.\label{force}
\end{equation}
The critical value of $\alpha$ at which $F=0$ is therefore
\begin{equation}
\alpha_c=\left|\frac{\partial C_{12}/\partial L}{\partial
C_{11}/\partial L}\right|-\sqrt{\left|\frac{\partial
C_{12}/\partial L}{\partial C_{11}/\partial
L}\right|^2-1}.\label{ac}
\end{equation}
When $\alpha<\alpha_c$, $F<0$, and vice versa. In
Fig.~\ref{fig:alpha}, we plot $\alpha_c$ as a function of $L/2R$
by taking the first 1000 terms in the series of Eq.~(\ref{pc}). We
see that the attraction is possible at any distance as far as the
ratio of the two potentials is smaller than $\alpha_c$. The closer
the two spheres are, the larger $\alpha_c$ is, and the easier the
attraction can be developed.

In order to gain more intuition about the interaction, also
determine the dependence of the attractive force on the distance
analytically, we study asymptotic behaviors of the force in the
case of $\phi_1>\phi_2>0$. In the limit of $L\gg 2R$, keeping the
leading order terms in Eq.~(\ref{c}), we get
\begin{equation}
C_{11}=DR+\frac{DR^3}{L^2}, \ \
C_{12}=-\frac{DR^2}{L},\label{pC12}
\end{equation}
and
\begin{equation}
F(L)=\frac{DR^2}{L^2}\phi_1\phi_2-\frac{DR^3}{L^3}(\phi_1^2+\phi_2^2).
\label{Fl}
\end{equation}
The physical meaning is clear. When the two spheres are far away
from each other, they interact like two point charges. In the zero
order, the magnitude of the charge is determined by their own
capacitance and potential, i.e., $DR\phi_1$ and $DR\phi_2$. This
gives the first term in Eq.~(\ref{Fl}) (one $D$ in charges is
cancelled out since the interaction itself is proportional to
$1/D$). In the next order, each point charge induces opposite
charge in the other sphere with the magnitude $-DR^2\phi/L$
following the standard method of images~\cite{Smythe}. The
interaction between the charge and its image gives the second term
in Eq.~(\ref{Fl}). From Eq.~(\ref{Fl}), one can easily get
$\alpha_c=R/L$. Therefore the attraction dominates only if
$\phi_2$ is much smaller than $\phi_1$. The attractive force is
proportional to $(R/L)^3$.

The more interesting limit is when the two spheres are very close
to each other, i.e., $d\ll R$. In this limit, we have
\begin{subequations}
\label{cd}
\begin{eqnarray}
C_{11}&=&\frac{DR}{4}\ln\frac{16\beta^2R}{d},\\
C_{12}&=&-\frac{DR}{4}\ln\frac{\beta^2R}{d},
\end{eqnarray}
\end{subequations}
where $\beta=1.78$ (see Appendix for the derivation of
Eq.~(\ref{cd})). According to Eq.~(\ref{Fs}), we have
\begin{equation}
F(d)=-\frac{DR}{8d}(\phi_1-\phi_2)^2.\label{fd}
\end{equation}
Remarkably, in this limit $\alpha_c=1$ and the force is always
attractive (the only exception is $\phi_1=\phi_2$ when the leading
order term given by Eq.~(\ref{fd}) is zero and one has to go to
the next order). Therefore we conclude that for any small
$\phi_1-\phi_2$, two conducting spheres always attract each other
at small enough distance. This originates from the fact that in
the limit of $d\rightarrow 0$, $C_{11}$ and $C_{12}$ both diverge
and diverge in the same way (see Eq.(\ref{cd})).

It should be mentioned that for the interaction between two
macroions, the model of conductors is valid only for $d\gg A$.
Indeed, instead of smeared charge distribution on a real conductor
surface, on the surface of the macroion, $Z$-ions are discrete and
form a WC-like liquid with ``lattice constant" $A$ (see
Fig.~\ref{fig:wc}). When $d$ becomes comparable with $A$, the
electric potential $\phi$ gets a periodic component along the
surface, which does not exist for a real conductor. In this paper,
we always focus on the case when $d\gg A$. Still, for $Ze\ll Q$
and, therefore, $A\ll R$, there is a big window of $A\ll d\ll R$
in which the model of conductors works and Eq.~(\ref{fd}) is
applicable.

\subsection{A small sphere close to a big sphere}

Now let us consider the interaction between a small sphere with
radius $R_1$ and a big sphere with radius $R_2$. We are interested
in the limit $R_1\ll R_2$ (Fig.~\ref{fig:spheres}b). We can solve
this problem by a similar procedure as in the last subsection,
starting from a formula like Eq.~(\ref{c}) but with both $R_1$ and
$R_2$ in it~\cite{Smythe}. Instead of doing this complicated
calculation, let us use physics intuition and look at the
interesting limit $d\ll R_2$. In this limit, the size of the large
sphere, $R_2$, becomes irrelevant to the problem, and the
interaction can be considered as the interaction between a sphere
with a metallic semi-space (see Fig.~\ref{fig:spheres}b).

Let us first consider the case when $\phi_2=0$. As well known in
electrostatics, a point charge induces an image charge in a
metallice semi-space which describes the interaction between the
charge and the metal. Similarly, using a series of image charges,
one can show that the interaction between a sphere and a metallic
semi-sphere is equivalent to the interaction between the sphere
and its ``image sphere" induced in the metal (see
Fig.~\ref{fig:spheres}b). The image sphere and the original sphere
have the same size, and their positions are symmetric about the
boundary of the metal. If the original sphere has fixed potential
$\phi_1$, its image sphere has fixed potential $-\phi_1$. Since
these two spheres are oppositely charged, we immediately see that
a conducting sphere is always attracted to a metallic semi-space.
Therefore spheres 1 and 2 attract each other.

Now we consider the case when $\phi_2\neq 0$. This potential is
equivalent to a charge sitting at the center of sphere 2, with
magnitude $DR_2\phi_2$. It induces an image charge in sphere 1. In
the limit of $R_2\rightarrow\infty$, this image charge is located
at the center of sphere 1 with the magnitude $-DR_1\phi_2$ (see
discussion of method of images in Ref.~\cite{Smythe}). It is
equivalent to add an potential $-\phi_2$ to sphere 1. As a result,
the potential of sphere 1 is renormalized to $\phi_1-\phi_2$.
Therefore, in this particular case, the interaction between two
spheres with potentials $\phi_1$ and $\phi_2$ is the same as two
spheres with potentials $\phi_1-\phi_2$ and 0. Clearly, the
attractive nature of the interaction between sphere 1 and its
image sphere is not changed by the nonzero $\phi_2$ (except the
special case $\phi_1=\phi_2$ when there is no interaction between
the two spheres).

Having established the attractive nature of the force, we now
calculate it quantitatively. Since the interaction is essentially
between sphere 1 and its image sphere, $C_{11}$, $C_{12}$ and
$C_{22}$ can be expressed through a linear combination of $C_{aa}$
and $C_{ab}$, where $C_{aa}$ and $C_{ab}$ are self and mutual
capacitances of sphere 1 and its image sphere given by
Eq.~(\ref{c}). We have
\begin{subequations}
\begin{eqnarray}
C_{11}(d)&=&C_{aa}(2d)-C_{ab}(2d),\\
C_{12}(d)&=&C_{ab}(2d)-C_{aa}(2d),\\
C_{22}(d)&=&DR_2+C_{aa}(2d)-C_{ab}(2d),
\end{eqnarray}
\end{subequations}
Here we put the $d$ dependence in each $C$ to remind us that the
distance between spheres 1 and 2 is $d$ but the distance between
sphere 1 and its image sphere is $2d$ (see
Fig.~\ref{fig:spheres}b). Noticing that $C_{11}=-C_{12}$ and
$\partial C_{11}/\partial d=\partial C_{22}/\partial d$, using
Eq.~(\ref{F}), we have
\begin{eqnarray}
F(d)&=&\frac{\partial C_{11}(d)}{\partial
d}\frac{(\phi_1-\phi_2)^2}{2}\nonumber\\ &=&-\left(\frac{\partial
C_{ab}(2d)}{\partial d}-\frac{\partial C_{aa}(2d)}{\partial
d}\right)\frac{(\phi_1-\phi_2)^2}{2}.
\end{eqnarray}
Since $\partial C_{aa}/\partial d<0$ and $\partial C_{ab}/\partial
d>0$ as we discussed in the last subsection, the force is
attractive for any given $\phi_1$ and $\phi_2$ (except the special
case $\phi_1=\phi_2$ when there is no interaction between the two
spheres).

The expressions of forces are particularly simple in certain
limits. When $R_1\ll d\ll R_2$, using Eq.~(\ref{pC12}), we have
\begin{equation}
F(d)=-\frac{DR_1^2}{4d^2}(\phi_1-\phi_2)^2.\label{Fd-}
\end{equation}
When $d\ll R_1$, using Eq.~(\ref{cd}), we have
\begin{equation}
F(d)=-\frac{DR_1}{4d}(\phi_1-\phi_2)^2.\label{fd2}
\end{equation}
This result can be compared with Eq.~(\ref{fd}) if $R_1=R$. We see
that the force is stronger by a factor of 2 in the case of two
spheres with different sizes. This is simply due to the fact that
the surfaces of the two spheres are ``closer" and the capacitances
diverge faster in the present case.

\section{The effect of screening by monovalent salt}
\label{sec:screening}

In a water solution of macroions and $Z$-ions, normally there is
also certain amount of monovalent salt. In most cases, the effect
of screening by monovalent salt can be described by the linear
Debye-H\"{u}ckel screening radius, $r_s$ (we discuss the
possibility and effect of nonlinear screening at the end of this
section). What we discussed in the last two sections corresponds
to the case $r_s\gg R_1, R_2, d$. Now we would like to consider
the opposite limit, $A\ll r_s\ll R_1,R_2$. We show that even
though attraction is suppressed and loses to repulsion at $d\gg
r_s$, the two spheres still attract each other at $d\lesssim r_s$.

\subsection{The method of images in the presence of monovalent salt}

Let us first discuss how the method of images is modified in the
presence of monovalent salt. We start from the simplest situation
in which there is a point charge $q$ close to a metallic
semi-space in a water solution with monovalent salt described by
screening radius $r_s$. When $r_s\rightarrow\infty$, it is well
known that the equal potential condition at the boundary of the
metal can be satisfied by putting an image charge $q^{\prime}=-q$
inside the metal at the position symmetric to $q$ (Here and below,
as a premise, water should also be introduced with the image
charge to produce the same dielectric constant). In the presence
of monovalent salt, one can easily check that the same boundary
condition can be satisfied by introducing the same image charge
$q^{\prime}$ at the same position, providing that a virtual cloud
of monovalent ions with the screening radius $r_s$ is produced
together with the image charge.

\begin{figure}[ht]
\begin{center}
\includegraphics[width=0.5\textwidth]{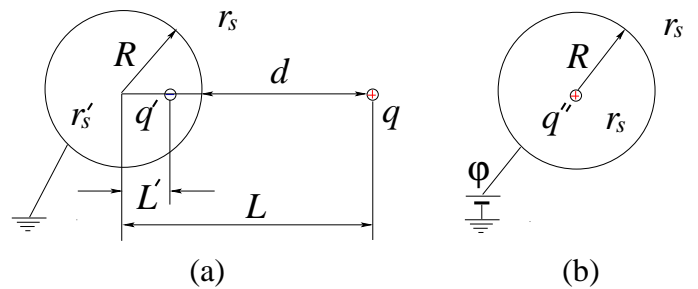}
\end{center}
\caption{Images in a conducting sphere in the presence of
monovalent salt with screening radius $r_s$: (a) a point charge
$q$ induces an image charge $q^{\prime}=-Rq/L$ with a virtual
cloud of monovalent ions with screening radius
$r_s^{\prime}=Rr_s/L$; (b) a conducting sphere with potential
$\phi$ is equivalent to a point charge $q^{\prime\prime}=DR\phi
e^{R/r_s}$ with a virtual cloud of monovalent ions with screening
radius $r_s$. }\label{fig:pointimage}
\end{figure}

As a more complicated situation, let us consider a point charge
$q$ and a grounded conducting sphere in a water solution with
monovalent salt described by $r_s$ (Fig.~\ref{fig:pointimage}a).
As well known~\cite{Landau}, when $r_s\rightarrow\infty$, the
magnitude and position of the image charge are given by (see
Fig.~\ref{fig:pointimage} for definition of all lengthes)
\begin{equation}
q^{\prime}=-q\frac{R}{L},\ \ L^{\prime}=\frac{R^2}{L}.\label{qL}
\end{equation}
When $r_s$ is finite, similar to the case of metallic semi-space,
a virtual cloud of monovalent ions should be created together with
the image charge. Due to the special geometry of a sphere, as one
can check, the screening radius of this virtual cloud is given by
\begin{equation}
r_s^{\prime}=\frac{Rr_s}{L},\label{rsp}
\end{equation}
different from $r_s$ in the solution.

Another relevant situation is a conducting sphere with fixed
potential $\phi$ in a water solution (see
Fig.~\ref{fig:pointimage}b). In this case, the potential in the
solution can be solved exactly using linearized Poisson-Boltzmann
equation and interpreted by the method of images. Actually it is
equivalent to the potential produced by a image charge
$q^{\prime\prime}$ at the center of the sphere. We can create a
virtual cloud of monovalent salt together with the image charge as
our convenience.  The magnitude of $q^{\prime\prime}$ is then
determined not only by the boundary condition but also the virtual
cloud we choose. As we will see in the next subsection, it is
convenient to introduce $r_s$ inside the sphere and get
\begin{equation}
q^{\prime\prime}=DR\phi e^{R/r_s}.\label{qprime}
\end{equation}

Let us calculate the interaction energy between a point charge and
a grounded conducting sphere (Fig.~\ref{fig:pointimage}a) for the
purpose of the next subsection. Considering the interaction energy
between $q$ and $q^{\prime}$, we have
\begin{eqnarray}
U&=&\frac{qq^{\prime}}{2D(L-L^{\prime})}e^{-(R-L^{\prime})/r_s^{\prime}}
e^{-(L-R)/r_s}\nonumber\\
&=&-\frac{Rq^2}{2D(L^2-R^2)}e^{-2d/r_s}, \label{Uqq}
\end{eqnarray}
where $d=L-R$. In the limit of $r_s\rightarrow\infty$, the
exponential factor goes to 1 and the energy goes back to the
standard expression. The additional factor $e^{-2d/r_s}$ can be
understood in the following way. The charge induced by $q$ on the
surface of the sphere is proportional to $e^{-d/r_s}$, the
interaction with it is also proportional to $e^{-d/r_s}$.

\subsection{Two spheres of the same size}

Now we consider the interaction between two conducting spheres of
the same size (see Fig.~\ref{fig:spheres}a) in the presence of
monovalent salt. Due to the complexity of the method of images in
the present case, instead of giving the general expression for the
force, we focus on two limits: $d\gg r_s$ and $d\ll r_s$.

Let us first discuss the case when $d\gg r_s$. As well known, when
$r_s\rightarrow\infty$, there is an infinite series of image
charges in each sphere~\cite{Smythe}. For finite $r_s$, similar to
what we discussed in the last subsection, the magnitudes and
positions of all image charges are the same but a virtual cloud of
monovalent salt is created together with each of them. In the
particular case of $d\gg r_s$, we can actually cut off the
infinite series and include the contribution from the leading
order image charges only. To see this, one just remember that the
image charges really represent surface charge densities induced on
the two spheres. Each image in the infinite series gives a
correction to the surface charge density calculated from the
previous order image. As we discussed after Eq.~(\ref{Uqq}), the
new correction induced gets one more factor $e^{-d/r_s}$ . Since
$d\gg r_s$, higher order corrections are exponentially small and
can be completely neglected.

\begin{figure}[ht]
\begin{center}
\includegraphics[width=0.45\textwidth]{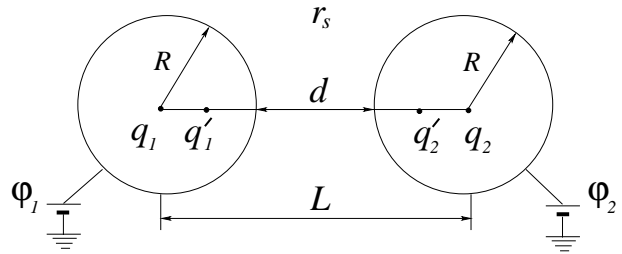}
\end{center}
\caption{The first two image charges induced in each sphere in
considering the interaction between two spheres. Each image charge
has its own virtual cloud (not shown).}\label{fig:sphereimage}
\end{figure}

In Sec.~\ref{sec:metal}, the energy is calculated using
capacitances (Eq.~(\ref{U})). In the present case, it is more
convenient to consider interaction energy between image charges
directly. The two methods are equivalent providing that a factor
$1/2$ is added to the interaction energy between a charge and its
image~\cite{Landau} in the later method (see Eq.~(\ref{UL})). The
two image charges in each sphere are given by (see
Fig.~\ref{fig:sphereimage})
\begin{subequations}
\label{q1122}
\begin{eqnarray}
q_1&=&DR\phi_1e^{R/r_s},\ \
q_2=DR\phi_2e^{R/r_s},\\
q_1^{\prime}&=&-\frac{DR^2\phi_2}{L}e^{R/r_s},\ \
q_2^{\prime}=-\frac{DR^2\phi_1}{L}e^{R/r_s}.
\end{eqnarray}
\end{subequations}
Here $q_1$ and $q_2$ are zero order image charges in each sphere
which take into account of $\phi_1$ and $\phi_2$ respectively,
similar to $q^{\prime\prime}$ discussed in the last subsection
(see Eq.~(\ref{qprime}) and Fig.~\ref{fig:pointimage}b). The
screening radius accompanied with them is just $r_s$. The first
order image charges $q_1^{\prime}$ and $q_2^{\prime}$ induced by
$q_2$ and $q_1$ are accompanied by $r_s^{\prime}=Rr_s/L$, similar
to $q^{\prime}$ discussed in the last subsection (see
Eqs.~(\ref{qL}), (\ref{rsp}) and Fig.~\ref{fig:pointimage}a).
Similar to Eq.~(\ref{Uqq}), the interaction energies between these
images are
\begin{eqnarray}
U(q_1,q_2)&=&\frac{DR^2\phi_1\phi_2}{L}e^{-d/r_s},\\
U(q_1,q_2^{\prime})&=&-\frac{DR^3\phi_1^2}{L^2-R^2}e^{-2d/r_s},\\
U(q_2,q_1^{\prime})&=&-\frac{DR^3\phi_2^2}{L^2-R^2}e^{-2d/r_s},
\end{eqnarray}
and
\begin{eqnarray}
U(L)&=&U(q_1,q_2)+\frac{1}{2}[U(q_1,q_2^{\prime})+U(q_2,q_1^{\prime})]
\nonumber\\
&=&\phi_1\phi_2\frac{DR^2}{L}e^{-d/r_s}
-\frac{\phi_1^2+\phi_2^2}{2}\frac{DR^3}{L^2-R^2}e^{-2d/r_s}.
\nonumber\\ \label{UL}
\end{eqnarray}
Correspondingly,
\begin{eqnarray}
F(L)&=&\phi_1\phi_2\frac{DR^2}{L}e^{-d/r_s}
\left(\frac{1}{r_s}+\frac{1}{L}\right)\nonumber\\
&-&(\phi_1^2+\phi_2^2)\frac{DR^3}{L^2-R^2}e^{-2d/r_s}
\left(\frac{1}{r_s}+\frac{L}{L^2-R^2}\right). \nonumber\\
\label{FLrs}
\end{eqnarray}
We see clearly that an additional factor of $e^{-d/r_s}$ appears
for each order of interaction. When $r_s\rightarrow\infty$, this
equation goes back to Eq.~(\ref{Fl}) as expected.

At $d\gg r_s$, the first term in Eq.~(\ref{FLrs}) dominates. When
$\phi_1=\phi_2$, we have the standard double layer repulsion of
DLVO theory~\cite{Colloid}. When $\phi_1$ and $\phi_2$ have same
signs, only the second term represents attraction, which is
negligible comparing with repulsion. This is analogous to
Eq.~(\ref{Fl}) but attraction is much weaker here. So in the limit
of $d\gg r_s$, the force is attractive only if potentials are
opposite in signs or one of potentials is equal to zero.

\begin{figure}[ht]
\begin{center}
\includegraphics[width=0.3\textwidth]{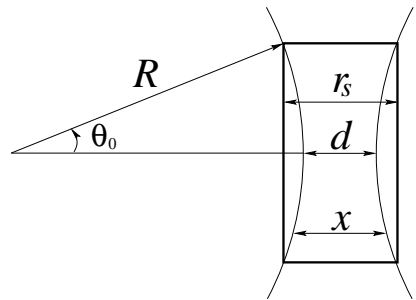}
\end{center}
\caption{A schematic illustration of the contact region between
two spheres. The contact region includes the surfaces of two
spheres inside a cylinder with the cross section shown by the
rectangular.}\label{fig:contact}
\end{figure}

Now let us consider the case when $d\ll r_s$. In this case, the
factor $e^{-d/r_s}\simeq 1$ and all higher order image charges
should be included. Since each image charge is accompanied by its
own virtual cloud, the interaction is very complicated. Instead of
calculating it exactly, we estimate $C_{11}$ and $C_{12}$ using a
simple method as follows. We divide the surface of the two spheres
into two pieces. In the first piece, the distance between the
surfaces of the two spheres is smaller than $r_s$ so that the two
spheres interact with each other in an unscreened Coulomb way. We
call this piece ``contact region" (see Fig.~\ref{fig:contact}). In
the second piece, the distance between the two surfaces is larger
than $r_s$ and the interaction between them is exponentially small
and negligible. Therefore, we can safely assume that the
modification to the capacitances of the two spheres due to their
proximity happens only in the contact region.

In order to calculate $C_{11}$ and $C_{12}$, let us first consider
a special case when $\phi_1=\phi_2=\phi$. In this case, the charge
induced on two spheres are equal and given by
$\phi(C_{11}+C_{12})$. In the contact region, the screening effect
by monovalent salt can be ignored and the adjacent spots of the
two spheres form a capacitor with zero voltage, therefore the
charge is zero~\cite{Rui}. While for the rest part of the sphere,
the interaction between two spheres is exponentially small and the
charge on each sphere is almost the same as if the other sphere is
not there. Therefore we have
\begin{equation}
C_{11}+C_{12}=
\frac{DR^2}{r_s}-\frac{DR(r_s-d)}{4r_s}.\label{C12+}
\end{equation}
Here $DR^2/r_s$ is the capacitance of a single sphere in the
presence of monovalent salt. When the sphere gets close to the
other sphere, it loses its capacitance in the contact region,
whose area is $\pi R(r_s-d)$.

Then we consider another special case when $\phi_1=-\phi_2=\phi$.
In this case, we have $q_1=(C_{11}-C_{12})\phi=-q_2$. Outside the
contact region, the capacitance is the same as in the previous
case. Inside the contact region, the capacitance can be estimated
as
\begin{equation}
C=D\int^{\theta_0}_0\frac{dS}{x(\theta)}
=D\int^{\theta_0}_0\frac{2\pi R^2sin\theta
d\theta}{d+2R(1-cos\theta)}=\pi DR\ln\frac{r_s}{d},
\end{equation}
where $\theta_0$ is the angle for the boundary of the contact
region (see Fig.~\ref{fig:contact}). The absolute value of the
charge induced on each sphere in the contact region is $2C\phi$.
Therefore
\begin{equation}
C_{11}-C_{12}=\frac{DR^2}{r_s}-\frac{DR(r_s-d)}{4r_s}+2\pi
DR\ln\frac{r_s}{d}.\label{C12-}
\end{equation}

Combining Eqs.~(\ref{C12+}) and~(\ref{C12-}), we get
\begin{subequations}
\label{Crs}
\begin{eqnarray}
C_{11}&=&\frac{DR^2}{r_s}-\frac{DR(r_s-d)}{4r_s}+\pi
DR\ln\frac{r_s}{d},\\
C_{12}&=&-\pi DR\ln\frac{r_s}{d}.
\end{eqnarray}
\end{subequations}
In the case of $r_s\sim R$, these results match Eq.~(\ref{cd})
which was obtained for $r_s\gg R$.

Using Eq.~(\ref{Fs}), and remembering $d\ll r_s$, we have
\begin{equation}
F=-\frac{\pi DR}{2d}(\phi_1-\phi_2)^2.\label{fds}
\end{equation}
The force is independent of $r_s$ in this limit because the change
of capacitances happens mainly in the contact region (the
logarithm term in Eq.~(\ref{Crs})). Again, the attraction shows up
at small $d$ and the attractive force decays as a power law. For
the model of conductors to work, we require $d\gg A$ which is
valid since $A\ll r_s$.

\subsection{A small sphere and a big sphere}\label{sec:sub}

Now we consider the interaction between a small sphere and a big
sphere (see Fig.~\ref{fig:spheres}b) in the presence of monovalent
salt. The approach we used is similar to that in the last
subsection.

When $d\gg r_s$, calculating the interaction between leading order
image charges similar to Eq.~(\ref{FLrs}), we get
\begin{eqnarray}
F(L)&=&\phi_1\phi_2\frac{DR_1R_2}{Lr_s}e^{-d/r_s}\nonumber\\
&-&\left(\frac{R_1\phi_1^2}{L^2-R_2^2}
+\frac{R_2\phi_2^2}{L^2-R_1^2}\right) \frac{DR_1R_2}{r_s}
e^{-2d/r_s}.\label{fLd}
\end{eqnarray}
Again, when $\phi_1$ and $\phi_2$ have the same signs, attraction
is exponentially weaker in this limit. When $d\ll r_s$, we can
estimate $C_{11}$, $C_{12}$ and $C_{22}$ similar to
Eq.~(\ref{Crs}). We get an attractive force given by
Eq.~(\ref{Fdrs}).
\\
\\
\\
In the end of this section, let us discuss the possibility of
nonlinear screening by monovalent salt in a special case when
$\phi_{1,2}>0$. The result can be generalized to other cases. The
condition for linear screening reads
\begin{equation}
\phi_{1,2}<\phi_c=\frac{k_BT}{e}|\ln(cv_0)|,\label{phic}
\end{equation}
where $e$ is the proton charge and $c$ and $v_0$ are the
concentration and volume of the monovalent ion. Here $\phi_c$ is
determined by entropy of monovalent ions in the solution.
According to Eqs.~(\ref{muc}) and~(\ref{phi12}),
$\phi_{1,2}\lesssim Ze/Dr_{1,2}\ll Ze/Da$, where $a\ll r_{1,2}$ is
the radius of a $Z$-ion (see Fig.~\ref{fig:wc}). If
$\phi_{1,2}>\phi_c$, $Ze/Da\gg \phi_c$. Therefore monovalent
counterions are adsorbed to $Z$-ions. As a result, the effective
charge of a $Z$-ion is renormalized to
\begin{equation}
Z^{\ast}e=Da\phi_c.
\end{equation}
This effective charge is not changed when a $Z$-ion is adsorbed on
the surface of the macroion since $Z^{\ast}e/Dr_{1,2}\ll
Z^{\ast}e/Da$. Consequently, $\phi_{1,2}$ should be calculated
using $Z^{\ast}$ and satisfy $\phi_{1,2}<Z^{\ast}e/Da=\phi_c$. In
summary, the nonlinear screening of monovalent salt is important
when $Z$-ions are strongly charged. In this case, our theory is
still applicable providing that $Z$ is replaced by $Z^{\ast}$
everywhere.

\section{attraction of two macroions, when only one of them adsorbs
$Z$-ions}\label{sec:experiment}

In this section, we generalize our theory to a little more
complicated case when the two bare macroions are oppositely
charged. As shown in Fig.~\ref{fig:afm}a, Positive $Z$-ions are
only adsorbed to the negative probe and invert its charge. In the
presence of large concentration of monovalent salt, monovalent
counterions are adsorbed to the positive surface (nonlinear
screening), reducing its electric potential to $\phi_c$ (see
Eq.~(\ref{phic})). This potential is certainly different from the
electric potential of the probe determined by $Z$-ions
(Eq.~(\ref{phi12})). We again have two conductors with different
fixed potentials. Even when both potentials are positive (the net
charges of the surface and the probe are positive), the force is
still attractive at small distances as we discussed in the last
section.

The atomic force experiment in Ref.~\cite{Lemay} is actually a
realization of the situation discussed above. The maximum of the
repulsive force mentioned in introduction can be understood
following Eq.~(\ref{fLd}), extrapolated to $d\simeq r_s$. When
both potentials are positive, at $d\gg r_s$, the first term
(repulsion term) in Eq.~(\ref{fLd}) dominates and the force is
repulsive. At $d\lesssim r_s$ ($r_s\simeq 100$ $\AA$ in the
experiment), the second, attraction term in Eq.~(\ref{fLd})
dominates due to its larger pre-factor of the exponential. This
leads to the maximum of the repulsion at $d \simeq r_s$. When
$d\ll r_s$, we have the polarization attraction given by
Eq.~(\ref{Fdrs}).

For $d<r_s$, Eq.~(\ref{ratio}) can be used to compare the
polarization attraction with the van der Waals one. According to
discussions in introduction and Sec.~\ref{sec:attraction}, we
conclude that for substantial difference between $\sigma_1$ and
$\sigma_2$, the polarization force $F_p$ is much larger than
$F_{vdW}$ at $d\gg A$ ($A=21$ $\AA$ for $Z=3$ and $-\sigma=-0.75$
e/nm$^2$). Therefore the van der Waals attraction has nothing to
do with the attraction observed in the experiment at $d\simeq 100$
$\AA$.

So far we have considered two cases when the electric potentials
of the two macroions are different and the attraction between
likely charged macroions is possible. One is that positive
$Z$-ions are adsorbed to both negative macroions; the other is
that positive $Z$-ions are only adsorbed to one negative macroion,
but monovalent counterions are adsorbed (nonlinearly screening) to
the other positive macroion. It is interesting to relate them to
the more standard case when two likely charged (say, negative)
macroions are nonlinearly screened by positive monovalent
counterions (no $Z$-ions in the solution). In this case, the two
macroions have the same electric potential $\phi_c$ given by
Eq.~(\ref{phic}) and never attract each other~\cite{Colloid}.

\section{Conclusion} \label{sec:conclusion}

In this paper, we discussed a long range polarization attraction
of two likely charged spherical macroions in the presence of
multivalent counterions ($Z$-ions). We show that the necessary
condition for the attraction is that the bare charge densities of
the two macroions are different. This polarization attraction is
much stronger and longer ranged than the van der Waals
force~\cite{Colloid} and the short range correlation
attraction~\cite{Bloomfield}. In the presence of large amount
monovalent salt, it adds an additional term to the standard double
layer repulsion of DLVO theory when the two macroions are
different. We discussed two cases when the polarization force
between two different macroions can be attractive, even if their
net charges have the same sign. In the first case, both macroions
adsorb $Z$-ions (Fig.~\ref{fig:afm}b). In the second case, one
macroion adsorbs $Z$-ions while the other adsorbs monovalent
counterions (Fig.~\ref{fig:afm}a). Here ``adsorb" means that the
binding energy is much larger than $k_BT$. In both cases, due to
different equilibrium conditions of adsorbed ions, the electric
potentials of two macroions are different and the attraction is
possible. On the other hand, the attraction is impossible if two
macroions are likely charged and both adsorb monovalent
counterions (no $Z$-ions in the solution). Our result
qualitatively agrees with atomic force
experiments~\cite{Lemay,Sivan}.

Even though in this paper we only discuss spherical macroions, the
polarization attraction we discovered can be generalized to other
geometries. Actually, as seen before, the attraction is always
developed at small distances between macroions when the overall
geometry is not very important. One can also understand this using
the language of contact electrification as discussed in the
introduction.

In this paper, in the connection with atomic force
experiments~\cite{Lemay,Sivan}, we focused on the force between
two macroions instead of the total free energy of the system.
Actually, for typical situations (e.g., $A\ll R$ or $A\ll r_s$),
one can show that the free energy may have the global minimum when
the two macroions are close to each other. Thus, the polarization
attraction can play an important role in determining the
equilibrium state of the system. It answers the gene delivery
related question: even if charges of DNA and cell membrane are
both inverted, they can still be attracted to each other since DNA
and membrane certainly have different surface charge densities. It
can also be important for aggregation or self-assembly of large
ensembles of different likely charged macroions with help of
oppositely charged $Z$-ions. Assume for example that we have a
mixture of equal numbers of two kinds of negative spheres with the
same radius but different values of the bare charges. Then in the
presence of positive $Z$-ions, they can attract each other and
assemble in a NaCl-like structure.
\appendix*
\section{}

Let us start from the asymptotic behavior of the series
$\sum_{n=1}^{\infty}1/\sinh nt$ at $t\rightarrow 0$. According to
the Cauchy Integral test of convergence, we have
\begin{equation}
\int_{1}^{\infty}\frac{1}{\sinh xt}dx
\leq\sum_{n=1}^{\infty}\frac{1}{\sinh
nt}\leq\int_{1}^{\infty}\frac{1}{\sinh xt}dx+\frac{1}{\sinh t}.
\end{equation}
Evaluating the integral, we have
\begin{equation}
\frac{1}{2t}\ln\frac{\cosh t+1}{\cosh
t-1}\leq\sum_{n=1}^{\infty}\frac{1}{\sinh
nt}\leq\frac{1}{2t}\ln\frac{\cosh t+1}{\cosh t-1}+\frac{1}{\sinh
t}.
\end{equation}
In the limit of $t\rightarrow 0$, it becomes
\begin{equation}
\frac{1}{t}\ln\frac{2}{t}<\sum_{n=1}^{\infty}\frac{1}{\sinh
nt}<\frac{1}{t}\ln\frac{2e}{t}.
\end{equation}
Therefore in this limit, we have
\begin{equation}
\sum_{n=1}^{\infty}\frac{1}{\sinh
nt}=\frac{1}{t}\ln\frac{2\beta}{t},
\end{equation}
where $\beta$ is a number satisfying $1<\beta<e$. Replacing $t$ by
$2t$, we get
\begin{equation}
\sum_{n=1}^{\infty}\frac{1}{\sinh
2nt}=\frac{1}{2t}\ln\frac{\beta}{t}.
\end{equation}
And
\begin{equation}
\sum_{n=1}^{\infty}\frac{1}{\sinh
(2n-1)t}=\sum_{n=1}^{\infty}\frac{1}{\sinh
nt}-\sum_{n=1}^{\infty}\frac{1}{\sinh
2nt}=\frac{1}{2t}\ln\frac{4\beta}{t}.
\end{equation}

Since $\cosh t=(2R+d)/2R$, the limit of $t\rightarrow 0$ is
equivalent to $d\rightarrow 0$ and $t=\sqrt{d/R}$ in this limit.
According to Eq.~(\ref{c}), we have
\begin{subequations}
\begin{eqnarray}
C_{11}&=&\frac{DR}{4}\ln\frac{16\beta^2R}{d},\\
C_{12}&=&-\frac{DR}{4}\ln\frac{\beta^2R}{d}.
\end{eqnarray}
\end{subequations}
Numerical calculation gives $\beta=1.78$.

\begin{acknowledgments}
The authors are grateful to R. Bruinsma, S. Lemay and U. Sivan for
useful discussions. This work was supported by NSF No.
DMI-0210844.
\end{acknowledgments}


\begin{thebibliography}{100}
\bibitem{potential} In this case, one can show that the polarization
attraction still exist, although its theory is more complicated.
\end{thebibliography}
\end{document}